\documentclass[aps,pra,twocolumn,nofootinbib,showpacs,floatfix]{revtex4-1}
\usepackage{bm}
\usepackage{graphicx}

\newcommand{\dens}[1]{{\rm [#1]}}

\newcommand{\tHe}{\mbox{$^3$He}}

\newcommand{\PRb}{P_{\rm Rb}}

\newcommand{\densrat}{{\cal D}}
\newcommand{\pinf}{P_{\infty}}

\newcommand{\be}{\begin{eqnarray}}
\newcommand{\ee}{\end{eqnarray}}

\begin{document}

 \title{Polarization Limits in K-Rb Spin-Exchange Mixtures}

\author{B. Lancor and T. G. Walker}
\affiliation{Department of Physics, University of Wisconsin-Madison, Madison, WI 53706}

\begin{abstract}
We present measurements of the optical absorption of K vapor at 795 nm due to the presence of high pressure He gas.   The results set a limit on the polarization attainable in hybrid spin-exchange optical pumping of \tHe.
\end{abstract}

\date{\today}

\pacs{32.70.-n,32.80.Xx,33.55.+b}

\maketitle

Hybrid spin-exchange optical pumping \cite{Babcock03} spin polarizes \tHe\ through spin-exchange collisions with an optically pumped mixture of K and Rb atoms. A dilute vapor of Rb is optically pumped in the usual manner \cite{WalkerRMP}, and polarizes a denser vapor of K by Rb-K spin-exchange collisions.   \tHe\ nuclei then become polarized through K-\tHe\ spin-exchange collisions.  The method takes advantage of the high efficiency of spin-exchange between K and \tHe\ \cite{Baranga98c}, while retaining the convenience of using 795 nm diode array bars for optical pumping of Rb.  Current state-of-the-art neutron spin-filters \cite{Boag09,Chen07b} and spin-polarized targets \cite{Riordan10,Ye10} utilize hybrid spin-exchange.

As long as the pumping light is sufficiently intense, the polarization achieved in hybrid spin-exchange experiments should be virtually the same as for single-species pumping.  However, several  experiments \cite{Babcock03,Chen07,Anger08,Singh10} have found that at high K/Rb density ratios $\densrat$ it is not possible to optically pump the alkali atoms to full polarization, even at very high optical pumping rates.   As proposed in \cite{Babcock03}, the most natural explanation for this is that there is weak off-resonant optical pumping of the K atoms by the 795 nm pumping light.  Assuming no spin-dependence to this rate, this acts as a light-induced spin relaxation mechanism that keeps the atoms from becoming fully polarized.  If the alkali-metal atoms are in  spin-temperature equilibrium, so that their electronic spin-polarizations $P$ are equal, and ground state spin relaxation can be ignored, the optical pumping equation becomes
\be
\dens{Rb}{dF_R\over dt}+\dens{K}{dF_K\over dt}=\dens{Rb}{R\over 2}(\pinf-P)-\dens{K}{R_K\over 2}P
\ee
which is basically a statement of angular momentum conservation.  The total angular momentum density $\dens{Rb}F_R+\dens{K}{F_K}$ of the Rb and K atoms increases by optical pumping of the Rb atoms at a rate $R$, increasing  $P$ towards its maximum possible value $\pinf$ \cite{Lancor10,*Lancor10b}.  The angular momentum is also lost by light absorption at a rate $R_K$ by the potassium atoms.  The factors of  1/2 assume relaxation of the electronic angular momentum in the excited state, and that sufficient N$_2$ quenching gas is included in the cell so that the nuclear spin is conserved in the excited-state \cite{Lancor10c}.

In steady-state, the polarization becomes
\be
P=\pinf{R\over R+\densrat R_K}\label{pss}
\ee
so the spin-polarization is significantly reduced when  $\densrat R_K$ becomes comparable to $R$.  Since both $R$ and $R_K$ are proportional to the pumping light intensity, the attainable polarization saturates at a value less than $\pinf$.  As has been noted before \cite{Babcock03,Lancor10}, the extreme optical depths of SEOP experiments make them particularly sensitive to such polarization limiting processes.

Using the apparatus described in Ref.~\cite{Lancor10}, we  have measured the absorption cross section $\sigma_K$ for K atoms near the Rb pumping wavelength of 795 nm.  We observed the transmission $1-e^{-n\sigma l}$ of a weak beam through 2 K cells, one $l=4.8$ cm diameter sphere containing 0.063 amg of N$_2$ and 2.93 amg of \tHe\, and an $l=5.7$ cm sphere with 0.083 amg N$_2$ and  0.924 amg of \tHe.  The transmission of a second probe beam at 855 nm, spatially overlapped with the first, was also monitored to account for any drifts in the cell transmissions that occur due to K droplet formation on the cell walls.  The potassium density $n$ and the helium density $\dens{\tHe}$ were deduced from K line-center absorption spectroscopy using the recently measured K-\tHe\ lineshape \cite{Singh10} with an assumed value for the line-broadening asymmetry parameter of zero.


\begin{figure}
\includegraphics[width=3.5 in]{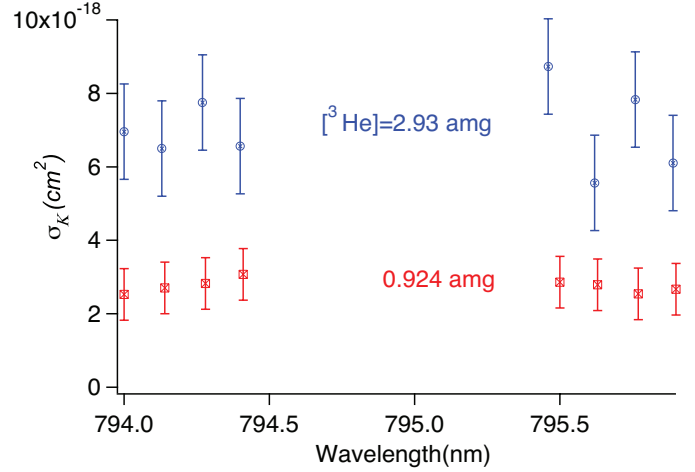}
\caption{Potassium absorption cross-section $\sigma_K$ as a function of wavelength in the vicinity of the Rb resonance at 795 nm, for two different \tHe\ densities.   } \label{fig:wlength}
\end{figure}Fig.~\ref{fig:wlength} shows the absorption cross-sections in the two cells found over a range of wavelengths near 795 nm. Due to  easily observable 1/10000 Rb contamination of these nominally pure K cells, we avoided measurements directly on the Rb resonance.  At the detunings used, Rb absorption was negligible.  At each wavelength, the absorption was measured at several temperatures, corresponding to several potassium densities.  A plot of the optical depth versus potassium density was made, and the cross section for absorption deduced from dividing the slope of a linear fit by the length $l$ of each cell.

The absorption cross section has no discernable frequency dependence, so the values at each wavelength were averaged to get the absorption cross section at 795 nm for each cell.
The ratio of the absorption cross sections  is, within uncertainty, equal to the ratio of the buffer gas densities in the two cells.  Since the \tHe\ densities greatly exceed the N$_2$ densities,  the dominant absorption process must be K-\tHe\ collisions.  Combining the results for the two cells we obtain the K-\tHe\ and K-N$_2$ cross sections
\be
\sigma_{\rm K-He}=2.19\pm .39 \times 10^{-18}{\mbox{cm$^2$}\over \mbox{amg}}\dens{\tHe}
\ee
and
\be
\sigma_{\rm K-N_2}=8.8\pm 7.6 \times 10^{-18}{\mbox{cm$^2$}\over \mbox{amg}}\dens{N_2}
\ee
The large uncertainty in the N$_2$ cross-section is due to its small abundance in the two cells. The uncertainty in the measurements arises mainly from etalon effects on the transmission of the 795 \ nm and 850 \ nm probe beams.  There are small contributions from uncertainty in the path length;  in the low \tHe\ density cell there is also a contribution from imperfect correction of transmission changes due to migration of K droplets on the cell wall.

The cross section being measured here corresponds to absorption in the quasistatic wings of the K resonance line\cite{Allard82}.  The spin-dependence of the absorption is discussed in our previous work \cite{Lancor10}.  Since the light is detuned by about 8 times the K fine-structure splitting, the absorption cross section is nearly spin-independent.  
Recent theoretical investigation of K-$^4$He far wing line broadening in the context of understanding the spectra of cool brown dwarfs \cite{Allard03} gives  $\sigma_{\rm K-^4He}=2.7 \times 10^{-18}{\mbox{cm$^2$}\over \mbox{amg}}\dens{^4He}$ at T=500K, within 2$\sigma$ of our result.


\begin{figure}[h]
\includegraphics[width=3.0 in]{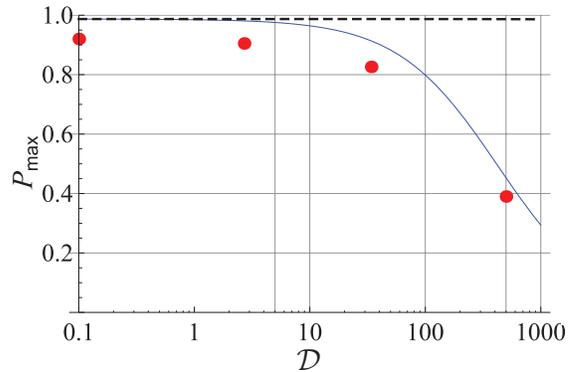}
\caption{Comparison of measured \cite{Babcock03} and modeled maximum achievable alkali polarization as a function of K/Rb density ratio $\densrat$ in a $\dens{\tHe}$=8.0 amg hybrid  cell pumped by a  1000 GHz bandwidth source. Modeling was done with (solid blue) and without (dashed black) off-resonant potassium absorption. }\label{fig:pmax}
\end{figure}

The observed cross section is in reasonable agreement with the value $R_K/R=2.2 \times 10^{-3}$ inferred by Babcock \emph{et al.} from the $\densrat$ dependence of $P_{\rm max}$, the alkali-metal polarization extrapolated to infinite pumping power\cite{Babcock03}.  Those results were obtained using an unnarrowed diode array bar as the pumping source for $\dens{\tHe}=8$ amg cells.  Solving  Eq.~\ref{pss} under those conditions with very large pump power, $\pinf$ from \cite{Lancor10,*Lancor10b} and our measured K-\tHe\ cross section, produces a curve in good agreement with the experiment at high $\densrat$ (Fig.~\ref{fig:pmax}).  (A linewidth of 1000 GHz was chosen for the pump laser, as an estimate of the unknown linewidth of the laser used in that experiment.)  Note that without potassium absorption, the model predicts high maximum alkali-metal polarization even at high $\densrat$.  At low $\densrat$, the observed polarizations were smaller than our simulations suggest, as discussed in detail previously \cite{Lancor10}.

\begin{figure}[b]
\includegraphics[width=3.5 in]{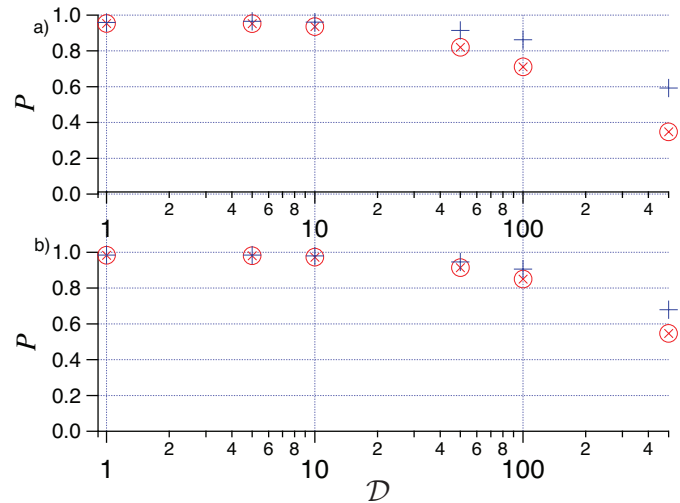}
\caption{Modeled average alkali polarization as a function of K/Rb density ratio $\densrat$ in a 7.9 cm long $\dens{\tHe}$=8.0 amg hybrid pumping cell at T=210$^\circ$C with (red) and without (blue) \ K-\tHe\ absorption, for a) a 100 W broadband  800GHz pumping source and b)  a 50 W narrowband 125 GHz pumping source. }\label{WideProp}
\end{figure}

The effect of K-\tHe\ absorption on optical pumping of $\dens{\tHe}$=8 amg cells with realistic pump power is shown in Fig.~\ref{WideProp}. The average alkali polarization under typical conditions is calculated with and without potassium absorption for a range of values of $\densrat$ using an optical pumping model that includes the effects of the ground state spin relaxation, spin relaxation at the cell walls, $\sigma_{\rm K-\tHe}$, $\pinf$, excited state nuclear spin relaxation \cite{Lancor10c}, ground state hyperfine splitting, and pump laser propagation.  For a 100 W broadband source at $\densrat$ above 5 the contribution from K-\tHe\ absorption becomes significant, and above $\densrat$=10 the alkali polarization quickly drops below 0.90.  It should be noted that the alkali polarization is limited to $\sim 0.95$ even at low $\densrat$, largely due to the  Rb-\tHe\ $\pinf$ effect.  With a 50 W narrowband source, the K-\tHe\ absorption noticeably reduces the alkali polarization, but $\PRb$ is above 0.90 up to a $\densrat$ of 50.

Alkali spin-polarization limits in high D cells of modest helium density(1-2 amg), pumped with narrowband ($\sim100$ GHz) laser sources, were reported in \cite{Chen07}.  Chen \emph{et al.} observed a decrease in the measured \tHe\ polarization with increasing $\densrat$ in three cells of $\densrat$=6.2, \ 46 and 155, and $\dens{\tHe}$=1.4,\ 1.9, and 1.1 amg respectively.  From measurements of the \tHe\ polarizations, they inferred alkali polarizations of only 0.77 in the $\densrat$=46 cell and 0.62 in the $\densrat$=155 cell.  There is an expectation that, with a fixed amount of laser power, the alkali-metal polarization should decrease with increasing $\densrat$ due to increased effective relaxation rate for the Rb atoms.  For the conditions described in \cite{Chen07}, our optical pumping simulations indicate this effect only accounts for a fraction of the observed drop in $P$ (Table ~\ref{PTab}).  However, potassium absorption also has little effect in narrowband pumping of low $\dens{\tHe}$ cells, and does not help explain the observed low alkali polarizations.  Although narrowband pumping gives better performance than broadband pumping, it is not as big an improvement as modeling would suggest \cite{Chen07}.
\begin{table}
\begin{tabular}{|l|l|l|l|}
\hline
$\densrat$ & $P(\rm Expt)$ & Theory, $R_K=0$ &  Theory, $R_K\ne 0$ \\ \hline
6.2        &  $\sim$.99   &   .987                &        .986            \\ \hline
46         & .77          &   .960                &    .954             \\ \hline
155        & .62          &  .838                 &    .825             \\ \hline

\end{tabular}
\caption{Comparison of measured \cite{Chen07} and predicted alkali polarizations $P$ in $\dens{\tHe}$ $\sim$ 1.5 amg cells of increasing $\densrat$. The low polarizations in high $\densrat$ cells at these helium densities are not explained by potassium absorption. }
\label{PTab}
\end{table}

Our measurement of the off-resonant pumping rate for K atoms explains the reduced performance of hybrid spin-exchange  optical pumping at high K/Rb ratios, but only for gas densities of several amagat.  Our simulations predict that the use of lower densities and narrower laser linewidths should greatly reduce off-resonant pumping effects.  Nevertheless,  it is  well-documented experimentally \cite{Chen07,Singh10} that the polarization still drops in high density ratio, low pressure hybrid cells pumped by narrowband light.  Thus there must be another as yet unknown mechanism at work, perhaps associated with the signficantly higher optical pumping rates in low pressure, narrowband experiments.

\acknowledgements{
This work was supported by the U.S. Department of Energy, Office of BasicÊ Energy Sciences, Division of Materials Sciences and Engineering, underÊ award DE-FG02-03ER46093, and aided by discussions with T. Gentile.
}

\bibliography{/Users/Thad_Walker/Research/thadbibtex/spinexchange}
\end{document}